# Metamaterial sound absorbers based on microperforated panels: an approach toward enhanced flexibility and near-limit broadband performance

*Jinjie Shi[1,†], Jie Luo[2,†,*], Chenkai Liu[1], Hongchen Chu[5], Yongxin Jing[1], Changqing Xu[5], Xiaozhou Liu[1,*], Jensen Li[3,4,*] and Yun Lai[1,*]*

[1]MOE Key Laboratory of Modern Acoustics, National Laboratory of Solid State Microstructures, School of Physics, Collaborative Innovation Center of Advanced Microstructures, and Jiangsu Physical Science Research Center, Nanjing University, Nanjing 210093, China

[2]School of Physical Science and Technology & Jiangsu Key Laboratory of Frontier Material Physics and Devices, Soochow University, Suzhou 215006, China

[3]Department of Physics, The Hong Kong University of Science and Technology, Clear Water Bay, Hong Kong, China

[4]Department of Engineering, University of Exeter, EX4 4QF, UK

[5]School of Physics and Technology, Nanjing Normal University, Nanjing 210023, China

†These authors contributed equally to this work

*Corresponding authors: Jie Luo (luojie@suda.edu.cn); Xiaozhou Liu (xzliu@nju.edu.cn); Jensen Li (jensenli@ust.hk); Yun Lai (laiyun@nju.edu.cn)

Abstract: Traditional microperforated panels (MPPs) and metamaterial-based sound absorbers rely on local resonances or multi-resonator designs, which limit their bandwidth, angular applicability, and ease of fabrication. Leveraging the reciprocity theorem and cavity resonances, we introduce a new class of robust MPP absorbers, termed meta-MPPs, capable of achieving ultrabroadband near-total sound absorption across a range of 0.37 to 10 kHz. These absorbers demonstrate average performance exceeding that of traditional MPPs by over 100%, approaching the theoretical causality limit. Notably, their absorption performance can be tuned between angularly asymmetric and omnidirectional modes and remains highly robust to variations in MPP parameters and geometrical configurations. Validated through simulations and experiments, our findings present a simpler, more robust, and highly adaptable solution for noise control.

## 1. Introduction

Sound absorption has long been a prominent research area in acoustics, with widespread applications in room acoustics and noise control engineering. Traditional porous materials [1], such as plastic foam, fiber glass, and mineral wool, have significant drawbacks such as poor dissipation at low frequencies, structural fragility, susceptibility to deformation, and difficulty in cleaning, etc. As a result, microperforated panels (MPPs) [2, 3] often combined with acoustic resonators, have emerged as a popular choice for indoor noise reduction. However, the bandwidth of conventional MPP sound absorbers remains significantly below the optimal performance predicted by the causality principle [4].

Over the past decades, acoustic metamaterials [5-14] have demonstrated remarkable capabilities in manipulating sound in unprecedented ways, leading to intriguing applications beyond the reach of traditional acoustic materials such as cloaking [15-19] and topological phenomena [20-26]. Metamaterial-based sound absorbers with subwavelength thickness have also been extensively explored, including decorated membranes [27, 28], air bubbles in rubber [29, 30], coiled-space structures [31-33], and combined Helmholtz resonators [34]. Recently, significant achievements have been made in broadening the operational frequency range, leveraging the resonant nature of metamaterials. These efforts include multiple resonances through optimized design [35-37] or coupled resonances [38, 39], coherent perfect absorption [40], and thermoacoustic energy conversion [41, 42]. It is exciting that some approaches have even approached the theoretical limit imposed by causality [35-37]. Nevertheless, the methods still require complex structure design or two-beam incidence, thus presenting challenges in fabrication and practical applications.

In this work, we present an approach to apply MPP techniques to metamaterials. The designed metamaterial MPPs, hereby termed as meta-MPPs, can flexibly allow either directional or omnidirectional sound absorption across an ultrabroadband spectrum. This meta-structure is composed of a tilted MPP array with sub-wavelength separation, as shown in Fig. 1a. This meta-MPP allows for near-reflectionless transmission under incidence at the tilt angle (left panel). Due to the reciprocity principle [43-47], reflectionless absorption is guaranteed at the opposite angle (middle panel). Since the mechanism does not rely on local resonances, the impedance-matched absorption spans an ultrabroad bandwidth, starting from the quasi-static limit. Such extreme angular asymmetry in transmission and absorption enables unprecedented applications, such as directional conversations (right panel). Notably, this asymmetric characteristic can be completely converted to symmetric by simply adding a wave-blocking backplate to the meta-MPP. Furthermore, by imposing appropriate sound-hard boundaries to the meta-MPPs to form open cavities, low frequency absorption can be enhanced through induced resonances. With an example of meta-MPP of 10 cm thickness, we demonstrate high absorption ($> 0.9$) across an ultrabroad frequency range ($0.37 \sim 10$ kHz) for a wide range of incident angles ($-75° \sim 75°$). Such an exceptional performance surpasses that of traditional MPP absorbers by over 100%, which achieves an average absorption of only 0.41 in the same frequency range. Furthermore, compared to other broadband metamaterial sound absorbers that rely on complicated resonant cavities [35-39], the meta-MPPs exhibit improved absorption characteristics with remarkable robustness on the MPP parameters and structural configurations. The robust and flexible absorption performance, with a bandwidth approaching the theoretical causality limit, indicates that our technique offers significant benefits for noise control.

## 2. Methods

### *2.1 Numerical simulations*

The full-wave simulations are performed using the commercial finite element software COMSOL Multiphysics. In the calculations, all solid structures are set as rigid, and the parameters of air are set as mass density $\rho_0 = 1.21\ \mathrm{kg/m^3}$, velocity $c_0 = 343\ \mathrm{m/s}$, dynamic viscosity $\eta = 1.81 \times 10^{-5}\ \mathrm{N \cdot S/m^2}$, and the preset environment temperature $T = 20\ °\mathrm{C}$. The background pressure field is used in Fig. 1b to f, 3d to f, 3g to i, 3k to m, 4c and d, 5b, d and f. The periodic boundary condition is set in the *x* and *y* direction and perfectly matched layers are adopted in the $z$ direction to reduce the reflection. In Fig. 4c and d, the surroundings of the absorber are set as sound-hard boundaries.

*2.2 Experimental measurements*

In Fig. 2a, the two scanning regions on the sides of reflection and transmission in the *x-y* plane are shown as two cyan rectangular areas, which are both of the sizes of $32 \times 30$ cm$^2$ and 3 cm away from the sample. A microphone (GRAS 46BE) is mounted on a moving stage to scan the acoustic field distribution at a step size of 10 mm. Quasi-Gaussian beam is generated by a speaker array (Five HiVi B1S) and a parabolic mirror. The experiment is carried out in an anechoic room to minimize reflection and noise. The measured reflected field distribution is obtained by subtracting the incident field (measured without the sample) from the total field (measured with the sample). All the measured field distributions in Fig. 2c and d have been normalized to the corresponding incident field. Sample I and sample II in Fig. 4 are fabricated by using etching technology and stereolithography three-dimensional printing techniques.

## 3. Results

*3.1. Theory and design of meta-MPPs*

Figure 1a illustrates the operational principle of the acoustic meta-MPP, comprising a periodic array of ultrathin MPPs tilted at an angle $\alpha$. The separation distance between adjacent MPPs is $d$, much larger than the thickness of each MPP $t_p$ (i.e., $d \gg t_p$), but significantly smaller than the wavelength $\lambda$ in air (i.e., $d < \lambda$). Initially, we consider a beam incident on the meta-MPP under an incident angle of $\theta_i = \alpha$ (left panel graph of Fig. 1a). Such waves propagate through the gaps between the MPPs, maintaining high transmittance and near-zero reflection due to the ultrathin MPPs, nearly independent of frequency (Fig. S1). An intriguing phenomenon occurs when the incident beam is flipped to the opposite angle, i.e., from $\alpha$ to $-\alpha$ (middle panel). Due to the system's reciprocity, near-zero reflection persists. However, transmission is significantly reduced as the waves squeeze through the perforations, resulting in substantial dissipation. By increasing the thickness of the meta-MPP, ultrabroadband perfect absorption for waves at $\theta_i = -\alpha$ can be attained. In short, ultrabroadband perfect transmission and perfect absorption coexist within the meta-MPP, albeit for different incident angles, i.e., $\theta_i = \alpha$ and $\theta_i = -\alpha$, revealing an extreme angular asymmetry. From another perspective, the ultrabroadband zero reflection at $\theta_i = \pm\alpha$ results from the non-resonant impedance matching between the meta-MPP and air.

The component of the meta-MPP, i.e., a single MPP, is characterized by a perforation diameter of $d_p = 0.1$ mm, a thickness of $t_p = 1$ mm, and an area fraction of $\sigma = 10\%$. The maximum absorptance of an MPP is 0.5 for wavelengths much larger than the MPPs' thickness[40] (Supplementary materials (note S1) and Fig. S2). Numerical validation is presented in Fig. 1b, showing the spectra of reflectance $R$, transmittance $T$, and absorptance $A$ for an MPP under normal incidence. The numerical calculation is performed by the commercial finite-element software COMSOL Multiphysics. As shown, the absorption remains below 0.5 across the studied frequency range of 1~6 kHz.

The meta-MPP is constructed as an array of such MPPs with a tilt angle $\alpha$. Figure 1c shows the absorptance of the meta-MPP as a function of the incident angle $\theta_i$ and the tilt angle $\alpha$ at 4 kHz. The thickness of the meta-MPP is set as $H = 7$ cm and the separation distance is set as $d = 3$ cm. Almost zero absorption is observed under $\theta_i = \alpha$ (white dotted lines), and high absorption ($A \geq 0.9$, black solid lines) is conversely achieved when $\theta_i = -\alpha$, matching perfectly with the previous analysis. To explore the performance of the meta-MPP, we set the tilt angle $\alpha = 45°$ and examine the reflectance (Fig. 1d), transmittance (Fig. S3), and absorptance (Fig. 1e) as functions of $\theta_i$ and frequency. Clearly, broadband and omnidirectional low reflection ($R < 0.1$) is observed for all angles $|\theta_i| \leq 70°$ and all frequencies below 6 kHz, beyond which diffraction effects become noticeable. This upper limit is determined by the separation distance between the MPPs (Supplementary materials (note S2) and Fig. S4). Moreover, broadband near-zero and near-total absorption are achieved around $\theta_i = 45°$ and $\theta_i = -45°$, respectively, consistent with our previous analysis. Figure 1f shows the calculated total acoustic field distributions of the meta-MPP at 4 kHz with $\theta_i = 45°$ and $\theta_i = -45°$, clearly demonstrating the asymmetric transmission and absorption, as well as the symmetric near-zero reflection. Figure 1g shows the calculated total acoustic field distributions of the meta-MPP under a point source. The meta-MPP leads to angle-selective radiation that creates a transparent window on the right side, but a blind region on the left side. Such broadband angle-selective radiation confirms the directional conversation in Fig. 1a.

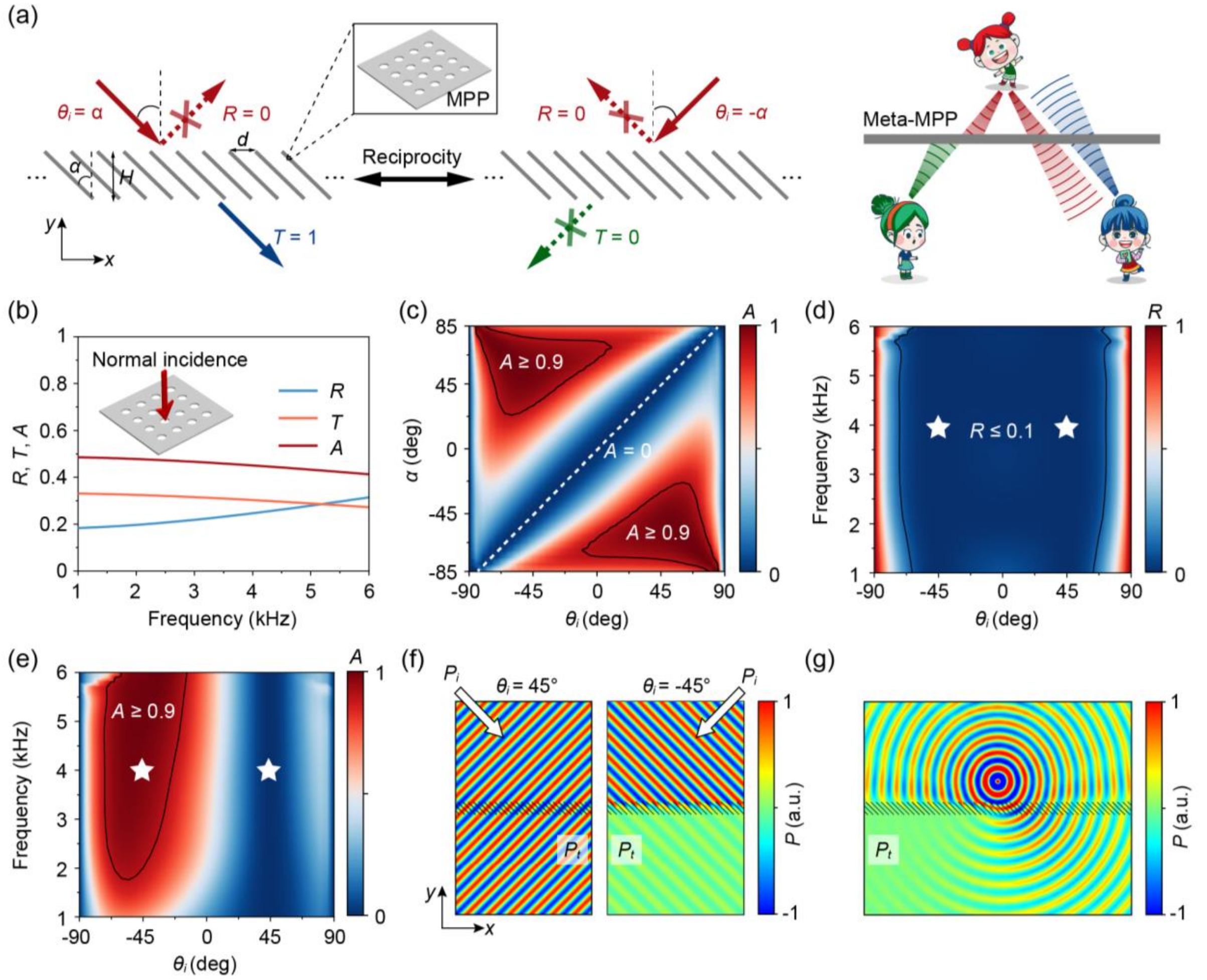

**Fig. 1.** The concept and principle of acoustic meta-MPPs for ultrabroadband absorption with extreme angular asymmetry. (a) Schematic diagram of the acoustic meta-MPP consisting of a periodic array of tilted ultrathin MPPs, which exhibits ultrabroadband reflectionless absorption and extreme angular asymmetry. (b) Spectra of reflectance $R$, transmittance $T$, and absorptance $A$ of a single MPP under normal incidence. (c) Absorptance of the meta-MPP as a function of the incident angle $\theta_i$ and tilted angle $\alpha$ of the MPP at 4 kHz. (d, e) Reflectance (d) and absorptance (e) of the meta-MPP regarding the incident angle and working frequency when $\alpha = 45°$. (f) Simulated total acoustic field distributions under incidence of plane waves at $\theta_i = 45°$ (left) and $\theta_i = -45°$ (right) at 4 kHz. (g) Simulated total acoustic field distribution under incidence from a point source located above the meta-MPP at 4 kHz.

*3.2. Experimental demonstration of meta-MPPs*

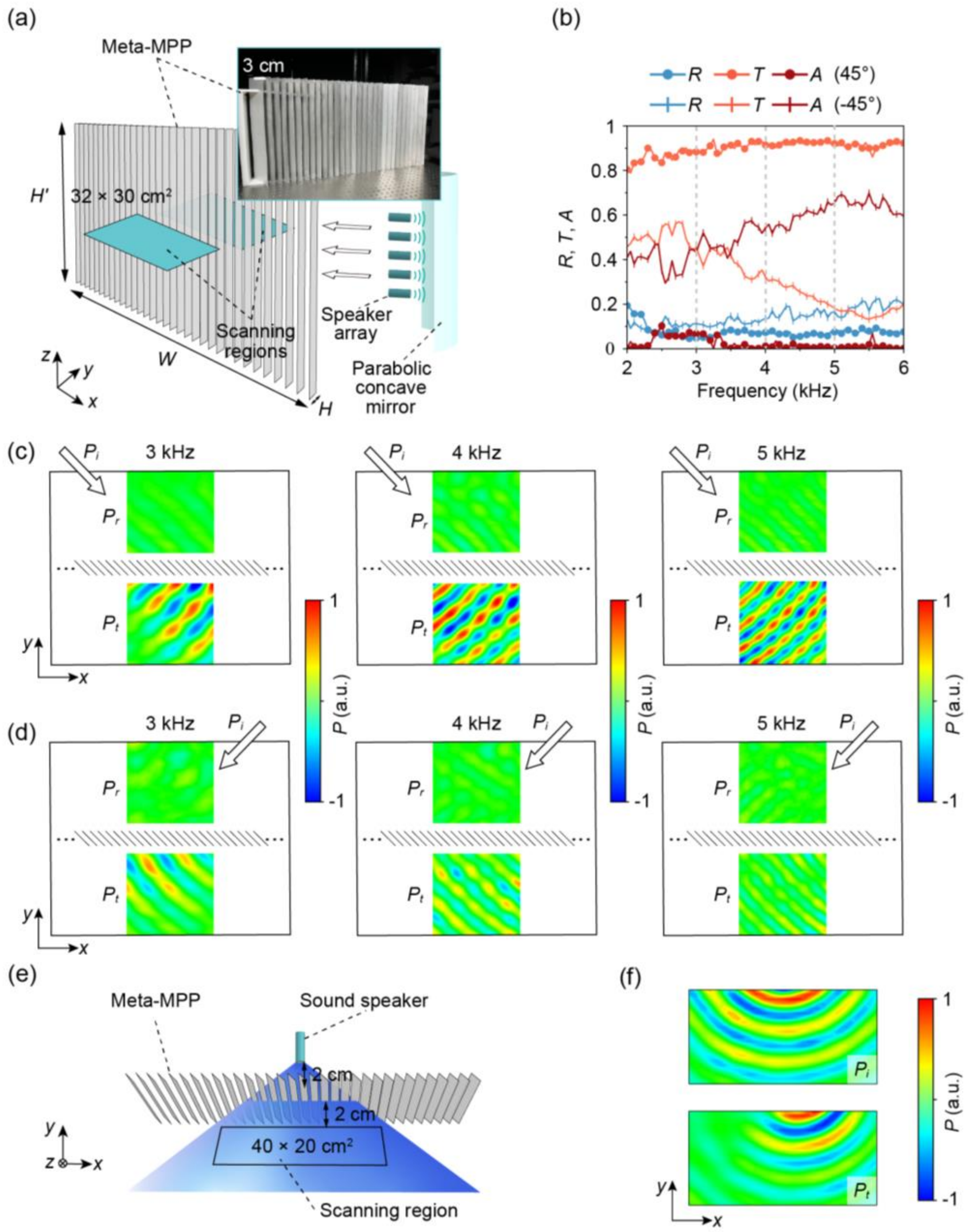

**Fig. 2.** Experimental demonstration of meta-MPPs. (a) Schematic diagram of the experimental configuration and the realization of a quasi-Gaussian beam. The inset shows the picture of the fabricated sample. (b) Measured reflectance $R$, transmittance $T$, and absorptance $A$ under incidence of the quasi-Gaussian beam at $\theta_i = \pm 45°$. (c, d) Measured pressure fields of transmitted (lower) and reflected (upper) waves under $\theta_i = 45°$ (c) and $\theta_i = -45°$ (d) at 3 kHz (left), 4 kHz (middle), and 5 kHz (right). (e) Schematic diagram of the experimental configuration under incidence from a point source. (f) Measured pressure fields without (upper) and with (lower) the meta-MPP at 5 kHz.

We now present the experimental validation of the exceptional broadband reflectionless absorption and extreme angular asymmetry enabled by the meta-MPP. Theoretical analysis shows that almost any kind of MPP can be employed for the construction of meta-MPP, almost regardless of its absorptance (Supplementary materials (note S3) and Fig. S5). Such a flexibility offers significant advantages in terms of ease of fabrication and robust performance. As a demonstration, a specific type of steel MPP with periodically arranged triangular cone holes is utilized here. The absorptance of this steel MPP is lower than 0.3 across the working frequency range (Fig. S6). A meta-MPP is assembled using 30 pieces of tilted steel MPPs ($\alpha = 45°$) with a separation distance of $d = 3$ cm, as shown in Fig. 2a. The meta-MPP has a thickness of $H = 7$ cm, a height of $H' = 50$ cm, and a width of $W = 100$ cm.

In the experiments, we first evaluate the spectra of reflectance $R$, transmittance $T$, and absorptance $A$ of the meta-MPP under incidence of the quasi-Gaussian beam at $\theta_i = \pm 45°$ (Fig. 2b). The source is constructed by a speaker array with a parabolic mirror, which can produce a quasi-Gaussian incident beam [48]. The transmittance $T$ is obtained as $|p_t/p_0|^2$, where $p_t$ and $p_0$ represent the pressure measured with and without the sample, respectively. Similarly, the reflectance $R$ is obtained as $|p_r/p_0|^2$. It is observed that the reflection is quite low for both cases of $\theta_i = \pm 45°$. However, the transmission and absorption exhibit distinct behaviors: high transmission under $\theta_i = 45°$ but high absorption under $\theta_i = -45°$, consistent with the theoretical predictions. We note that the absorptance exceeds the 0.5 limit, but it is not notably high due to the relatively low dissipation of each MPP and the relatively thin thickness of the meta-MPP. Figure 2c and d shows the measured acoustic field distributions of the transmitted (lower) and reflected (upper) waves under $\theta_i = \pm 45°$ at 3 kHz, 4 kHz, and 5 kHz, respectively. The scanning regions are highlighted in cyan in Fig. 2a. These results, again, validate the broadband low-reflection property as well as the angular asymmetric behavior.

We have also investigated the performance of the meta-MPP under point source incidence (Fig. 2e). The scanning region is 2 cm away from the meta-MPP, as marked by the black region ($40 \times 20$ cm$^2$) in Fig. 2e. Figure 2f presents the measured pressure fields without (upper) and with (lower) the meta-MPP at 5 kHz. A clear angular asymmetry in transmission is observed, creating a blind region on the left side but a transparent region on the right, in agreement with the results shown in Fig. 1g.

*3.3. Meta-MPPs for ultrabroadband and omnidirectional absorption*

In most scenarios of sound absorption, omnidirectional absorption is generally preferred over angle-dependent absorption. In the following, we demonstrate that by modifying the boundaries of meta-MPPs, the original directional absorption can be converted into omnidirectional near-total absorption with an extremely broad bandwidth, approaching the theoretical limit predicted by the causality principle.

We begin with a two-dimensional meta-MPP (model I, Fig. 3a), with $\alpha = 60°$, $d = 1$ cm, $L = 10$ cm, and $H = 10$ cm. The relevant MPP parameters are $d_p = 0.2$ mm, $t_p = 0.1$ mm, and $\sigma = 2\%$. Periodic boundaries are applied in the *x* direction, and an open boundary is set for the bottom. In this scenario, model I is equivalent to the meta-MPP studied in Fig. 1. Figure 3d presents the simulated absorptance of model I as a function of the incident angle and frequency. Angular asymmetry is observed in absorption, where near-zero absorption occurs at $\theta_i = \alpha = 60°$ and near-perfect absorption at $\theta_i = -\alpha = -60°$, consistent with previous analysis. Subsequently, we change the bottom-open boundary to a bottom-closed boundary, i.e., adding a wave-blocking backplate (model II, Fig. 3b). Interestingly, this change completely eliminates the angular asymmetry and significantly enhances the absorption performance (Fig. 3e). This improvement arises because the transmitted waves at $\theta_i = -\alpha = -60°$ are completely reflected by the sound-hard boundary and then largely absorbed by the meta-MPP. High absorption ($A > 0.9$) is achieved for all angles $|\theta_i| \leq 75°$ and spans an ultra-wide frequency range (0.68∼16.8 kHz at $\theta_i = \pm 60°$). Nevertheless, the absorptance within the incident angle range of $-60°$∼$60°$ at low frequencies (below 1.5 kHz) is less than 0.9 (the green dotted region in Fig. 3e), indicating a need for further improvement. Figure 3f provides a magnified view of this region.

Remarkably, the absorption performance in the low-frequency regime can be further enhanced by changing the periodic boundaries in the $x$ direction to sound-hard boundaries (model III, Fig. 3c). The thickness of the hard wall is set to be $t = 2$ mm, with a gap of $s = 5$ mm between the left wall and the meta-MPP. This gap allows sound to penetrate into the meta-MPP, further improving absorption. A detailed discussion can be found in Supplementary materials (note S4) and Fig. S7. Figure 3g displays the simulated absorptance of model III under normal incidence, in comparison with the results obtained for model II and an MPP with a 9.8 cm-thick cavity (Fig. S8). The grey region indicates the additional absorption of model III compared to model II, showing a significant enhancement in the low frequency regime. To elucidate the underlying physics, the normalized sound intensity (color map) and velocity fields (black arrows) for models II and III at 0.33 kHz are presented in Fig. 3h and i, respectively, which correspond to the points $P_1$ and $P_2$ in Fig. 3g. The sound intensity is uniformly distributed in model II, but it becomes highly non-uniform in model III. The non-uniformity indicates that cavity resonances occur within the hard box, thereby enhancing the absorption performance. The curvatures of the velocity stream (indicated by black arrows) of model III demonstrate stronger coupling among the MPP channels than that of model II.

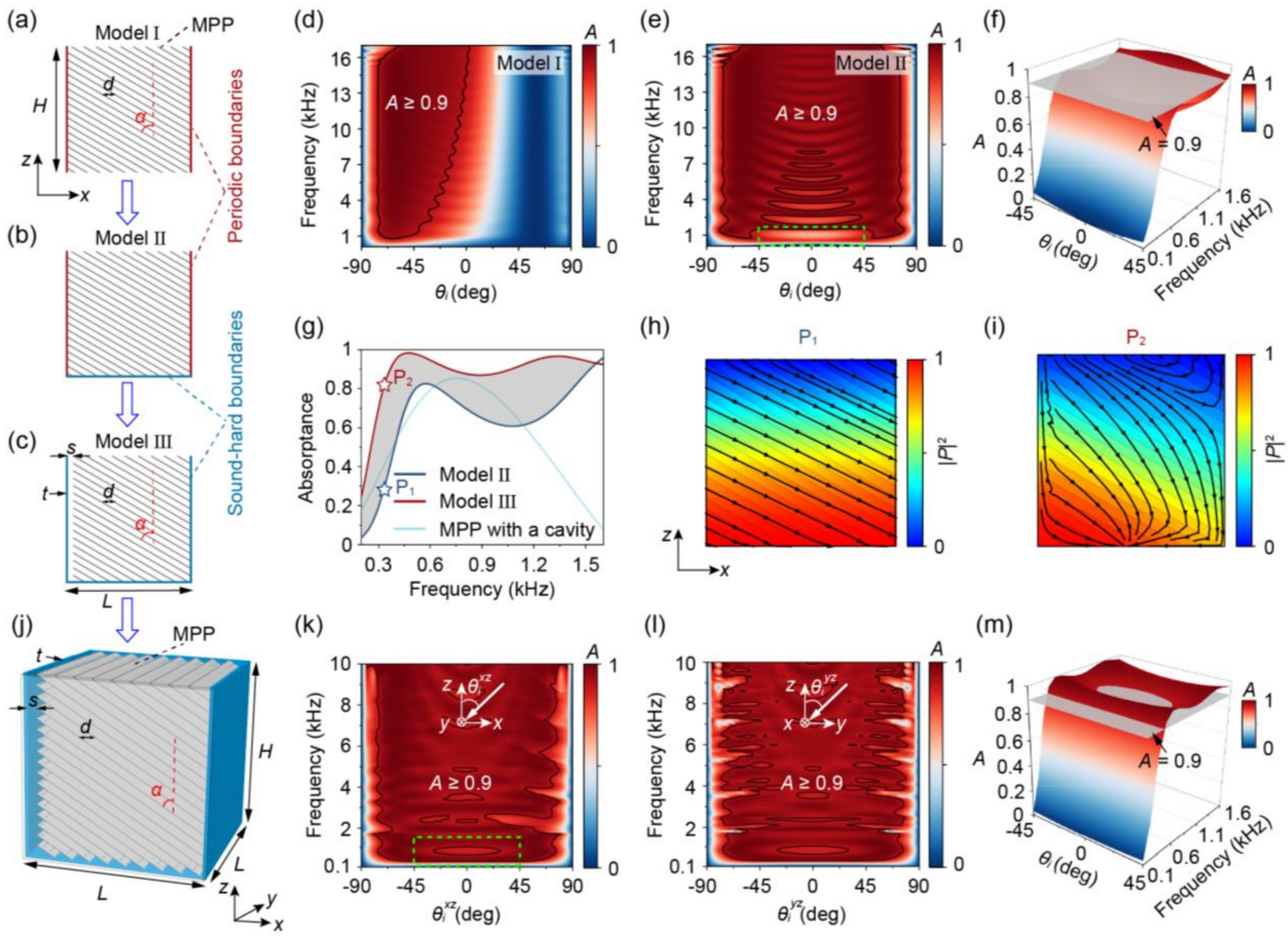

**Fig. 3.** Design of meta-MPPs as ultrabroadband omnidirectional sound absorber. (a-c) Schematic diagrams of two-dimensional models with a bottom-open boundary (model I), with a bottom-closed boundary (model II), and with a bottom-closed boundary and sound-hard boundaries in the *x* direction (model III). (d, e) Absorptance of model I (d) and model II (e) as functions of the incident angle and the frequency. (f) A magnified view of the absorptance of model II, corresponding to the green dotted region in (e). (g) Absorptance of model II, model III, and MPP with a 9.8 cm-thick cavity under normal incidence. (h, i) Simulated normalized intensity (color map) and velocity field (black arrows) at 0.33 kHz. The two graphs correspond to the points $P_1$ and $P_2$ in (g). (j) Illustration of a three-dimensional sound absorber constructed by adding sound-hard boundaries in the *y* directions of model III. (k, l) Absorptance of the three-dimensional absorber as a function of the incident angle $\theta_i^{xz}$ (k) and $\theta_i^{yz}$ (l) and the frequency. (m) A magnified view of the absorptance of the three-dimensional sound absorber, corresponding to the green dotted region in (k).

We then extend the model III to three dimensions through adding sound-hard boundaries along the *y* direction to form a three-dimensional open cavity (Fig. 3j). The front panel is set transparent to reveal the internal structures. Figure 3k shows the simulated absorptance of the three-dimensional sound absorber as a function of the incident angle $\theta_i^{xz}$ and the frequency for incident wave vector within the *x*-*z* plane. High absorption ($A > 0.9$) is observed within the range of angles $-75°\sim75°$. When $\theta_i^{xz} = 0$ (or $\theta_i^{xz} = -\alpha = -60°$), the average absorptance is 0.92 (or 0.97) over the frequency range of $0.37\sim10$ kHz. The thickness of the sound absorber is only 5% exceeding the minimum value predicted by causality (Supplementary materials (note S5)). Similar excellent absorption performance is also observed for incident wave vector within the *y*-*z* plane (Fig. 3l). To explicitly demonstrate the enhanced absorption performance at low frequencies by modifying the boundaries, Fig. 3m provides a magnified view of the absorption of the absorber (corresponding to the green dotted region in Fig. 3k). Compared to the previous two-dimensional absorber (Fig. 3f), this practical three-dimensional absorber demonstrates significantly better low-frequency absorption performance.

It is noteworthy that selecting an appropriate tilt angle for the MPPs is important in achieving the ultrabroadband omnidirectional absorption (Fig. S9). A larger tilt angle $\alpha$ ensures the ultrabroadband impedance matching at larger incident angles, i.e., $|\theta_i| = \alpha$, thereby facilitating high absorption at these angles. It is expected that if all MPPs are arranged vertically (i.e., $\alpha = 0$), the absorption performance deteriorates significantly (Supplementary materials (note S6) and Fig. S10). The tilt of the MPP arrays plays an important role in absorption.

*3.4 Experimental demonstration of meta-MPP absorbers*

To validate the ultrabroadband omnidirectional sound absorber, we have conducted measurements using the impedance tube method (Fig. 4a). This method allows accurate measurement of absorptance, but is limited to normal incidence. Figure 4b displays a photograph of two fabricated samples, labeled sample I and sample II. Sample I (or II) is larger (or smaller), intended for evaluating absorption performance at lower (or higher) frequencies. Details of the smaller sample are presented in Fig. S11. In the measurements, two square impedance tubes (BSWA F100 and BSWA F50) with side lengths of 10 cm and 5 cm, respectively, were utilized. Figure 4c shows the simulated absorption of sample I (dark red lines) and sample II (light red lines) under normal incidence, demonstrating significantly higher absorption in the frequency range of 0.37∼10 kHz compared to an MPP with a 9.8 cm-thick back cavity (light blue lines). The results confirm that such meta-MPPs can enhance the average absorptance of traditional MPP absorbers from 0.41 to 0.92 across the frequency range of 0.37∼10 kHz. Figure 4d provides a magnified view of the grey area in Fig. 4c. The experimental measurements (dots) agree well with the simulation results (lines), confirming the outstanding sound absorption performance of the fabricated samples. It should be noted that the measured frequency range is limited to the range of 0.2∼3.3 kHz due to the limitation of the impedance tubes. A square impedance tube with a side length of 10 cm (or 5 cm) has a cutoff frequency of approximately 1.6 kHz (or 3.3 kHz), beyond which the measured results are inaccurate. Although the experiments were conducted under normal incidence, the meta-MPP absorber remains effective under oblique incidence, as shown in Fig. S12.

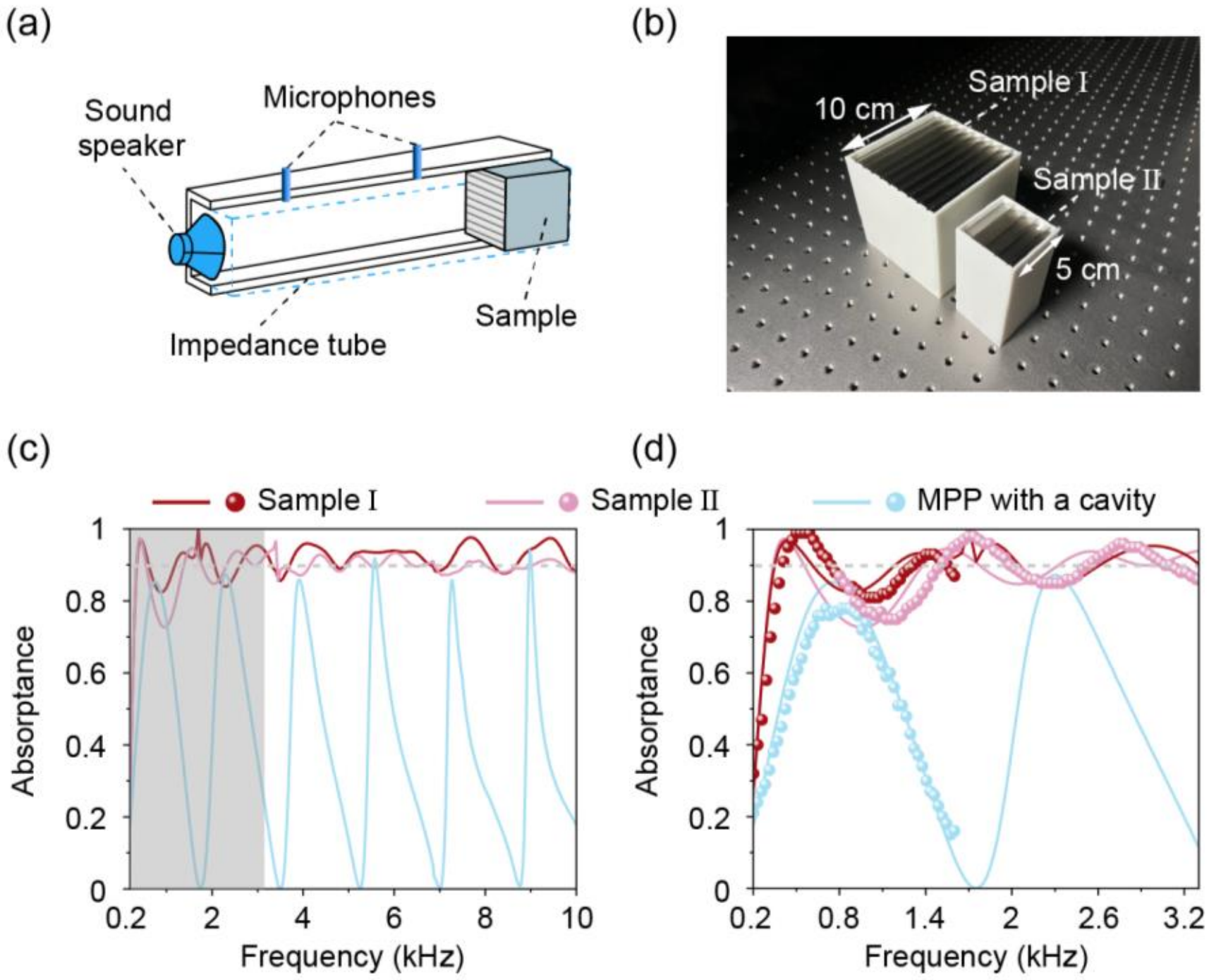

**Fig. 4.** Experimental demonstration of meta-MPPs as ultrabroadband omnidirectional absorbers. (a) Schematic diagram of the experimental setup of the impedance tube. (b) Photograph of sample I and sample II. (c) Simulated absorption spectra of sample I, sample II, and an MPP with a 9.8 cm-thick back-cavity. (d) A magnified view of the grey region in (c). The lines and dots denote simulated and measured results, respectively.

*3.5 Robustness of meta-MPP absorbers*

Notably, the designed meta-MPP absorbers exhibit exceptional robustness, as variations in structural configurations and MPP parameters within a certain range have negligible effects on their absorption performance. This robustness is illustrated through three representative scenarios.

In the first scenario, the MPPs in the meta-MPP absorber from Fig. 3j become randomly distributed, while all other parameters remain unchanged. The schematic cross-section in the *x*-*z* plane is shown in Fig. 5a, and the corresponding absorptance as a function of the incident angle $\theta_i^{xz}$ and the frequency is depicted in Fig. 5b. It is evident that the random distribution of MPPs has minimal impact on the absorption performance of the meta-MPP absorber, which consistently maintains its ultrabroadband and omnidirectional absorption capabilities. In the second scenario, the tilt angle of the meta-MPP absorber from Fig. 3j is increased to $\alpha = 70°$, with all other parameters unchanged. The schematic cross-section in the *x*-*z* plane is presented in Fig. 5c, and the corresponding absorption performance is shown in Fig. 5d. The results indicate that the meta-MPP absorber retains its excellent ultrabroadband and omnidirectional absorption characteristics despite the tilt modification. In the third scenario, the MPPs in the meta-MPP absorber from Fig. 3j are replaced with seven types of MPPs with random parameters, as shown in Fig. 5e. Despite these modifications, Fig. 5f reveals that the meta-MPP absorber maintains excellent absorption performance. The detailed parameters of the meta-MPP absorbers can be found in Fig. S13. The remarkable robustness of the designed meta-MPP absorbers is attributed to their non-resonant nature and reciprocity. This property is critical for practical fabrication and long-term application, as it significantly reduces manufacturing complexity and minimizes performance degradation caused by aging or external factors. Unlike other broadband metamaterial absorbers [35-39] that rely on intricate resonant cavities, the meta-MPP absorbers offer a simpler and more robust solution.

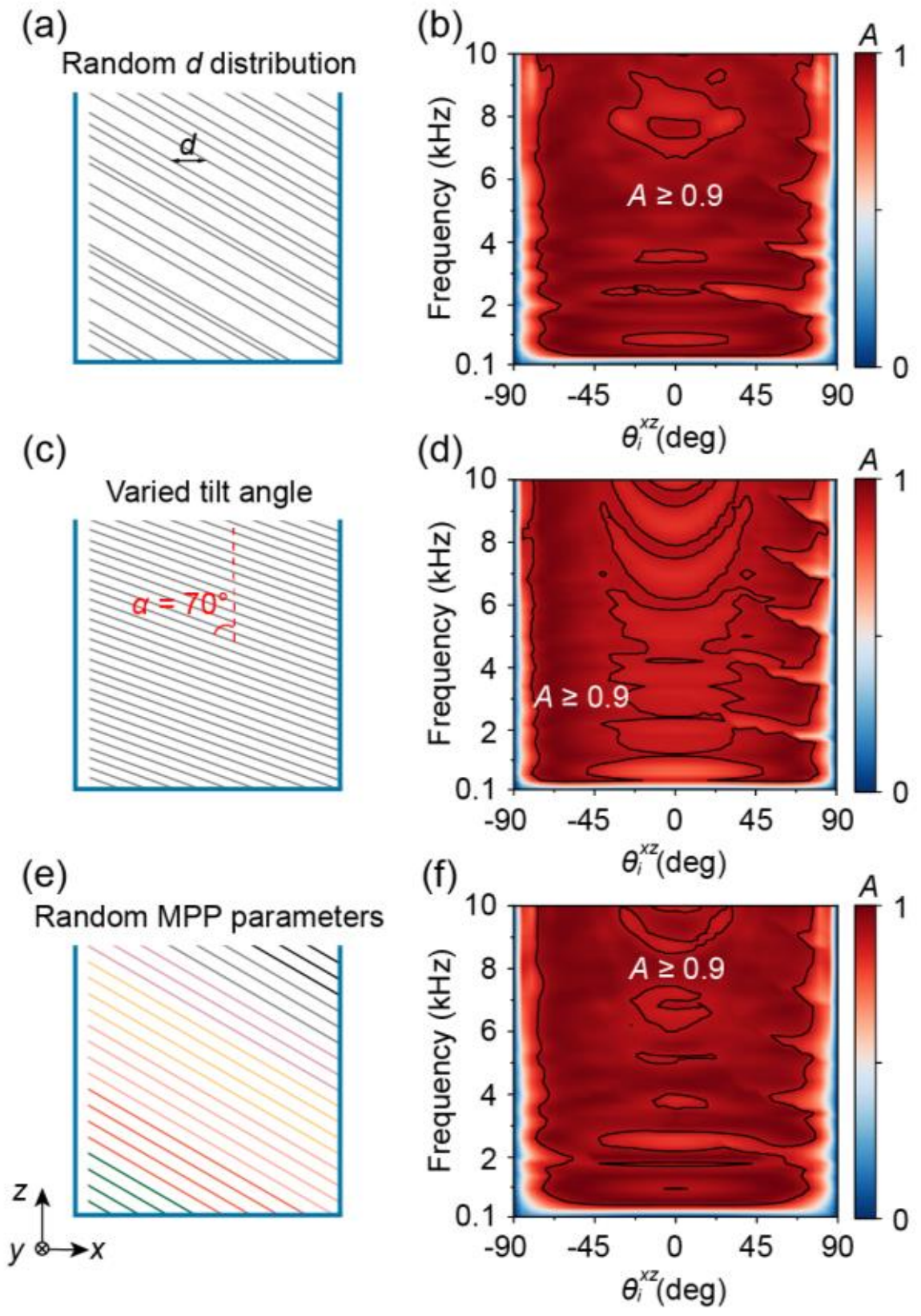

**Fig. 5.** Robustness demonstration of meta-MPP absorbers. (a-f) Schematic cross-sections of the meta-MPP absorber in the *x*-*z* plane and the corresponding absorptance as a function of the incident angle $\theta_i^{xz}$ and the frequency under three scenarios: random *d* distribution (a, b), varied tilt angle (c, d), and random MPP parameters (e, f).

## 4. Discussion and conclusion

The meta-MPPs proposed here have the salient advantages of extremely large working bandwidth and flexible switching between directional and omnidirectional absorption. The bandwidth of reflectionless absorption can, in principle, cover an ultra-broad spectrum starting from the quasi-static limit, far surpassing the capabilities of traditional MPP absorbers. Attributed to the design based on the reciprocity theorem as well as the simple conditions on forming cavity resonances, the meta-MPPs exhibit remarkable robustness against the structural or material parameters of the unit, along with the attractive advantages of easy fabrication and applicability for a wide range of incident angles.

The design principle of meta-MPPs is very general. Any type of MPP with a substantial amount of dissipation can be employed for constructing meta-MPPs, such as textiles and foam metals. Even for MPPs with relatively weak dissipation, meta-MPPs with ultra-broadband and omnidirectional near-total absorption can still be constructed, at the cost of a relatively larger thickness. In other words, the impedance matching between meta-MPPs and air is almost independent of the detailed choice of MPPs as well as the frequency. Therefore, meta-MPPs can easily improve the absorption performance beyond traditional MPPs, without resorting to complicated resonance design [35-39].

In conclusion, we present a straightforward yet highly effective approach to achieving ultrabroadband and omnidirectional absorption using tilted MPP arrays, termed meta-MPPs. These meta-MPPs exhibit extreme angular asymmetry, capable of switching between perfect transparency and perfect absorption upon reversing the incident angle. By imposing a hard backplate and boundaries to the meta-MPPs, such angular dependence can be eliminated, eventually achieving omnidirectional and ultrabroadband near-perfect absorption, approaching the causality limit. Our approach offers a profound shift in metamaterial sound absorbers, paving the way for the next generation of high-performance, flexible, robust sound absorbers.

**Conflict of Interest**

The authors declare that they have no conflict of interest.

**Acknowledgments**

Yun Lai thanks the support from the National Natural Science Foundation of China (12474293 and 12174188) and the Natural Science Foundation of Jiangsu Province (BK20233001). Xiaozhou Liu acknowledges the support from the National Key R&D Program of China (2020YFA0211400), the National Natural Science Foundation of China (12174192), and the State Key Program of National Natural Science Foundation of China (11834008). Jie Luo thanks the support from the National Natural Science Foundation of China (12374293) and the Natural Science Foundation of Jiangsu Province (BK20221354). Jensen Li acknowledges the support from the support from the Research Grants Council of Hong Kong (R6015-18). Jinjie Shi thanks the support from the China Postdoctoral Science Foundation (2023M731612). Hongchen Chu acknowledges the support from the National Natural Science Foundation of China (12404364) and the Natural Science Foundation of Jiangsu Province (BK20240575).

**Author Contributions**

Yun Lai and Jie Luo conceived the idea. Jinjie Shi conducted the analysis and simulations. Chenkai Liu provided support on experimental design and sample fabrication. Jinjie Shi conducted the experiments. Yongxin Jing and Chenkai Liu helped in the experiments. Hongchen Chu and Changqing Xu helped in the theoretical analysis. Yun Lai, Jie Luo, Jensen

Li, and Xiaozhou Liu organized and led the project. All the authors contributed to the data analysis and manuscript preparation.

## Appendix A. Supplementary materials

Supplementary materials to this article can be found online at.

# Appendix A. Supplementary materials

**Note S1 Dependence of the absorptance on different geometrical parameters of the MPP**

We have investigated the dependence of the absorptance on different geometrical parameters of the MPP under normal incidence, as shown in Fig. S2 (a to c). When one geometric parameter is varying, all the other parameters are set as fixed. Fig. S2a illustrates the simulated absorptance regarding the frequency and perforation diameter. Clearly, the absorptance is maximum when the perforation diameter is close to 0.1 mm, which proves that the perforation diameter plays a significant role in the absorption of MPP. Next, the plate thickness and area fraction are considered in Fig. S2b and c. It is seen that the MPP exhibits superior absorption capabilities when the plate thickness is less than 5 mm and the area fraction is less than 0.2. These results can be employed to select appropriate parameters for MPP design. Finally, Fig. S2d displays the simulated absorptance of the MPP as a function of area fraction and perforation diameter, with a plate thickness of 1 mm and a working frequency of 4 kHz. In all figures, the white dotted lines and star denote the selected parameters in our theoretical design. It is noted that the maximum absorptance of each MPP is 0.5 for wavelengths much larger than the MPPs' thickness.

On the other hand, the impedance of MPP can be well approximated by the following formula [14]:

$$Z_p = -i\omega \frac{\rho_0 t_p}{\sigma}\psi_1 + \frac{32\eta t_p}{\sigma {d_p}^2}\psi_2 \tag{S1}$$

where $\psi_1 = 1 + \frac{1}{\sqrt{1+\frac{\xi^2}{2}}} + 0.85\frac{d_p}{t_p}$, $\psi_2 = \sqrt{1+\frac{\xi^2}{32}} + \frac{\sqrt{2}}{32}\xi\frac{d_p}{t_p}$, and $\xi = d_p\sqrt{\omega\rho_0/(4\eta)}$. $\rho_0$ and $\eta$ refer to the static air density and dynamic viscosity of air.

In the case of a plane wave at normal incidence to the MPP, the transmission and reflection coefficient are written as

$$T = \left|\frac{2Z_0}{Z_p+2Z_0}\right|^2 \text{ and } R = \left|\frac{Z_p}{Z_p+2Z_0}\right|^2 \tag{S2}$$

the absorption coefficient for an incident wave on one side is

$$A = 1 - T - R = \frac{1}{2}\left(1 - \left|\frac{Z_p - 2Z_0}{Z_p+2Z_0}\right|^2\right) \tag{S3}$$

The resistance $Z_p = 2Z_0$ maximizes the absorption coefficient of the MPP, i.e., $A = \frac{1}{2}$, which matches our simulation results.

**Note S2 Discussion on the diffractions of the meta-MPPs**

According to Bragg's Law, the diffraction frequency $f_m$ can be determined by the following relationship:

$$f_m = \frac{mc_0}{2d sin\theta_i} \tag{S4}$$

where $d$ represents the periodic length of the structure, $\theta_i$ is the incident angle, $m$ denotes the diffraction order, and $c_0$ is the speed of sound. It is noteworthy that $f_m$ typically decreases as the incident angle increases. For the scenario discussed in the main text ($d = 3$ cm), considering first-order diffraction ($m = \pm 1$), the diffraction frequency is $f_m = 5736.57$ Hz when $\theta_i = \pm 85°$. On the other hand, the diffraction frequency can be elevated by reducing the separation distance between the MPPs, i.e., $d$. Fig. S4a and b shows the simulated absorptance of the meta-MPP regarding the incident angle and working frequency for $d = 2$ cm and $d = 1$ cm, respectively. All other parameters remain consistent with those in Fig. 1e. According to Equation (S4), for $d = 2$ cm, the diffraction frequency is $f_m = 8618.09$ Hz when $\theta_i = \pm 85°$. For $d = 1$ cm, the diffraction frequency is $f_m = 17216.89$ Hz when $\theta_i = \pm 85°$. The simulation results, as depicted in Fig. S4a and b, clearly align with the theoretical predictions.

**Note S3 Robustness of the meta-MPPs**

We emphasize that almost any kind of MPP can be used to construct a meta-MPP, almost regardless of the absorption. For example, we employ MPPs with relatively low absorptance to create a meta-MPP similar to the design shown in Fig. 1e. This meta-MPP has a thickness of $H = 7$ cm. The relevant MPP parameters are $d_p = 0.15$ mm, $t_p = 1.8$ mm, and $\sigma = 20\%$. Fig. S5a plots the simulated absorptance of the meta-MPP regarding the incident angle and working frequency. It achieves broadband near-zero absorption around $\theta_i = 45°$, while partial absorption occurs around $\theta_i = -45°$. This is due to the relatively low dissipation of each MPP. Notably, by simply increasing the thickness of the meta-MPP, ultrabroadband perfect absorption for waves under $\theta_i = -45°$ can be obtained. Fig. S5b shows the simulated absorptance of the meta-MPP regarding the incident angle and working frequency when $H = 14$ cm. Clearly, the bandwidth of absorption around $\theta_i = -45°$ is significantly broadened, consistent with our analysis. Despite the increased thickness ($H = 14$ cm), the meta-MPP still exhibits extreme angular asymmetry, similar to that observed with $H = 7$ cm in Fig. S5a. On the other hand, we plot the reflectance of the two cases, as shown in Fig. S5c and d, respectively. Both cases exhibit broadband and omnidirectional low reflection, demonstrating that this characteristic is independent of the meta-MPP's thickness. Therefore, the meta-MPP can be constructed of any kind of MPP, showcasing its remarkable robustness.

**Note S4 Simulated absorptance curves of model III for different sizes of an opening, s**

To demonstrate the impact of an opening on the absorption of model III, we first calculate the absorption curve of model IV (without an opening), as depicted in Fig. S7b. The schematic diagram of model IV is illustrated in Fig. S7a. Clearly, compared to model IV, model III ($s =$ 5 mm) exhibits superior absorption between 0.3 kHz to 0.8 kHz. The grey region indicates that the absorption of model III ($s = 5$ mm) is greater than that of model IV, which can help to evaluate the improved performance. Additionally, we plot the absorption curves for various opening sizes within the same figure. In comparison to model III with $s = 3$ mm, model III with $s = 5$mm demonstrates a notable improvement in maximum absorption. Furthermore, compared to model III with $s = 7$ mm, model III with $s = 5$ mm shows enhanced absorption at lower frequencies. Consequently, model III with $s = 5$ mm is deemed the optimal choice here.

**Note S5 Causality constraint**

Material response function for the incident wave must satisfy the causality principle, which leads to an inequality that relates a given absorption spectrum to the sample thickness [35]:

$$H \geq \frac{1}{4\pi^2}\frac{B_{eff}}{B_0}\left|\int_0^\infty ln[1 - A(\lambda)]\, d\lambda\right| = H_{min} \tag{S5}$$

where $\lambda$ and $A(\lambda)$ represent the sound wavelength in air and absorption coefficient, respectively. $B_0$ is the bulk modulus of air, and $B_{eff}$ denotes the effective bulk modulus of the sound-absorbing structure in the static limit. By inserting the simulated absorption for normal incidence in Fig. 3k into Equation (S5), we obtain $H_{min} = 9.52$ cm, which is close to the thickness (10 cm) of the designed omnidirectional sound absorber. Therefore, based on theoretical design, we conclude that our omnidirectional sound absorber has approached the limit of causality.

**Note S6 Simulated absorption performance of the absorber when all MPPs are arranged vertically**

To demonstrate how reciprocity enhances absorption, we investigate the absorption performance of an absorber with vertically arranged MPPs. The schematic diagram of the absorber in the *x*-*z* plane is provided in Fig. S10a. Here, the relevant parameters are $d = 1$ cm and $s = 5$ mm. Fig. S10b and c, illustrate, respectively, the simulated absorptance of the absorber regarding the incident angle $\theta_i^{xz}$ and $\theta_i^{yz}$ and working frequency. At normal incidence when the incident wave vector lies on the *x*-*z* plane, zero absorption is observed. However, as the incident angle increases, the absorber starts to demonstrate some absorption, although it remains relatively low overall. Conversely, when the incident wave vector lies on the *y*-*z* plane, there is almost no absorption. In this scenario, the incident beam aligns parallel to the surface of each MPP, preventing the waves from interacting with the MPPs. These results underscore the significance of reciprocity in augmenting absorption.

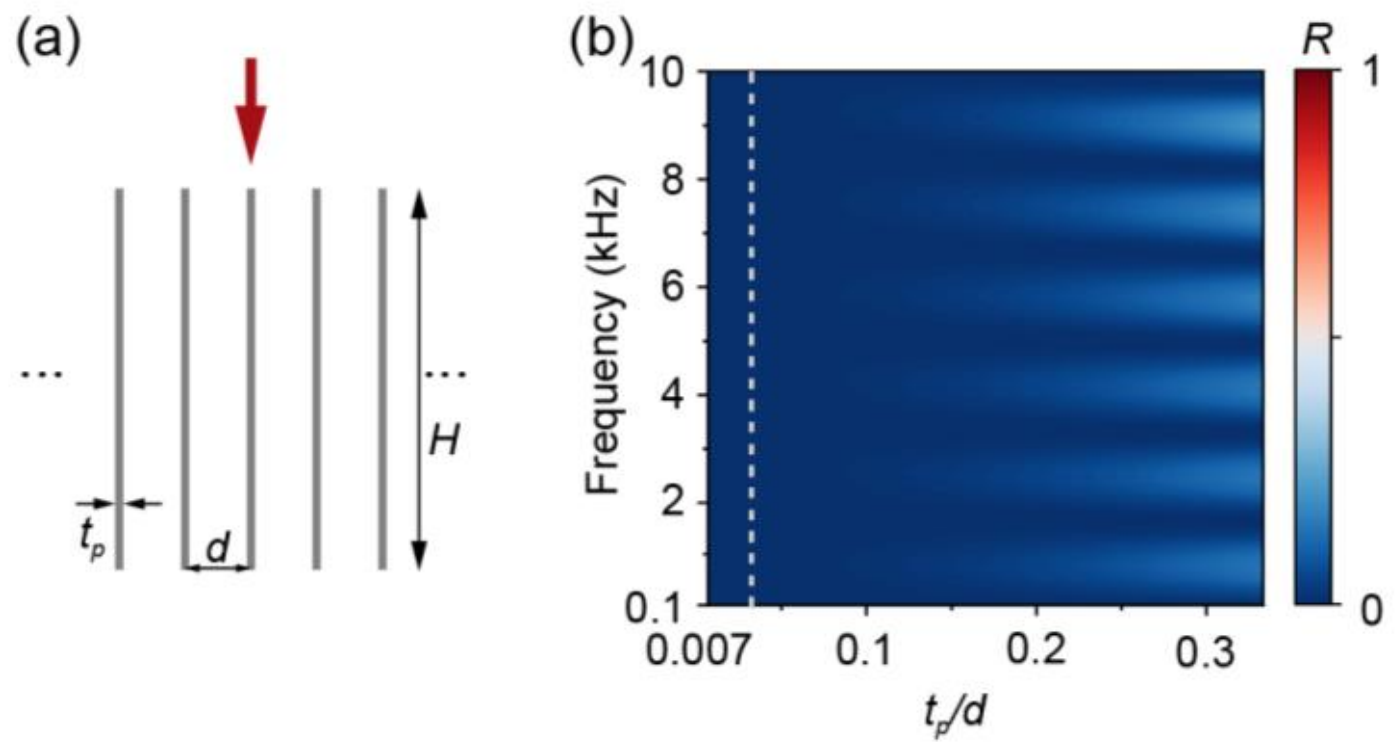

**Fig. S1.** Reflectance of the MPP array when sound waves are incident along the MPP array. (a) Schematic diagram of the vertically arranged MPP arrays, where the thickness of the MPP array is $H = 10$ cm. (b) Reflectance of the MPP array regarding $t_p/d$ and working frequency under normal incidence. Notably, the reflectance of the MPP array becomes negligible when the thickness of each MPP $t_p$ is sufficiently small. In Fig. 1a, $t_p/d$ is 0.033, corresponding to the grey dotted line in Fig. S1b.

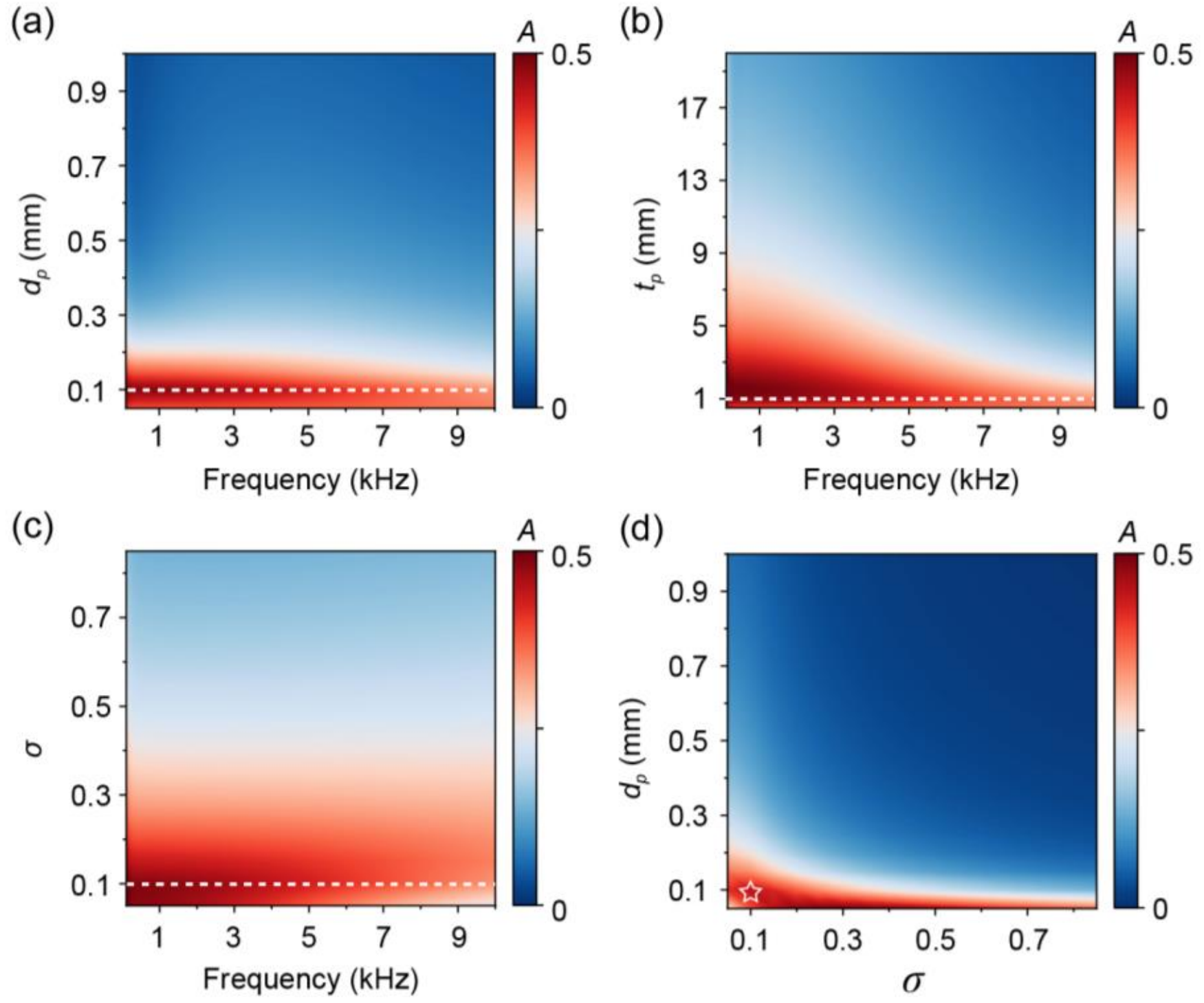

**Fig. S2.** Dependence of the absorption properties on different geometrical parameters of the MPP. (a-c) Simulated absorptance curves of the MPP varying with the perforation diameter, $d_p$ (a), plate thickness, $t_p$ (b), and area fraction $\sigma$ (c). (d) Simulated absorptance curves of the MPP as a function of the area fraction, $\sigma$, and perforation diameter, $d_p$. White dotted lines and the white star in the colormaps represent the selected parameters in our design.

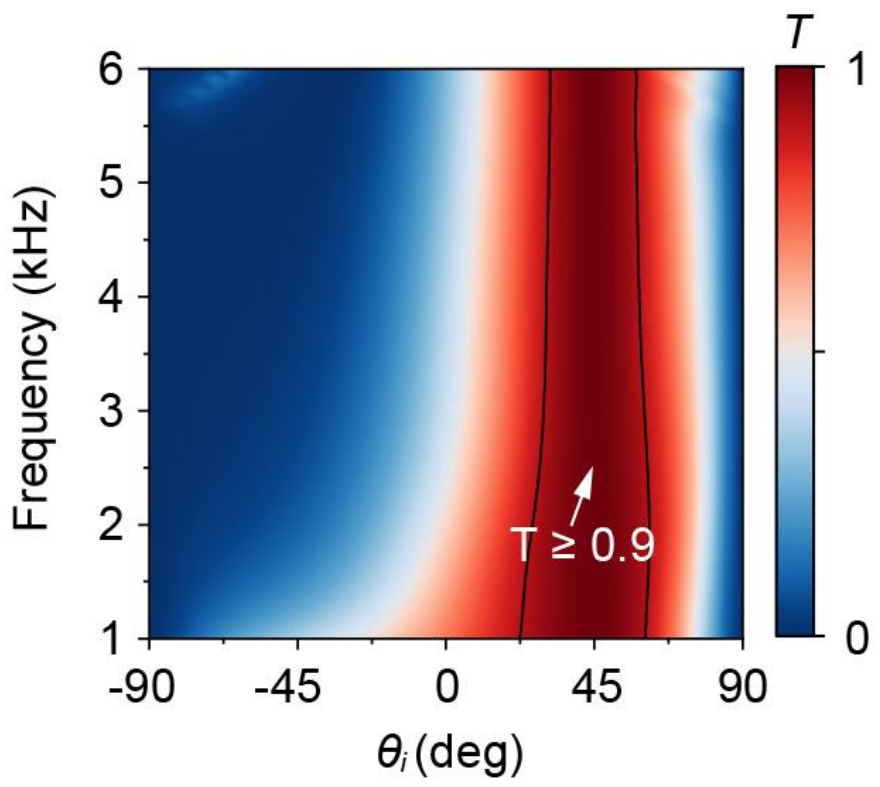

**Fig. S3.** Simulated transmittance of the meta-MPP. It is seen that the transmission exhibits a markedly asymmetric behavior. The perfect transmittance occurs under $\theta_i = 45°$ over a broad spectrum. Particularly, when the incident angle deviates away from 45°, the transmission is still quite high, indicating some angular insensitivity in the transmission performance, which will correlate with the absorption behavior.

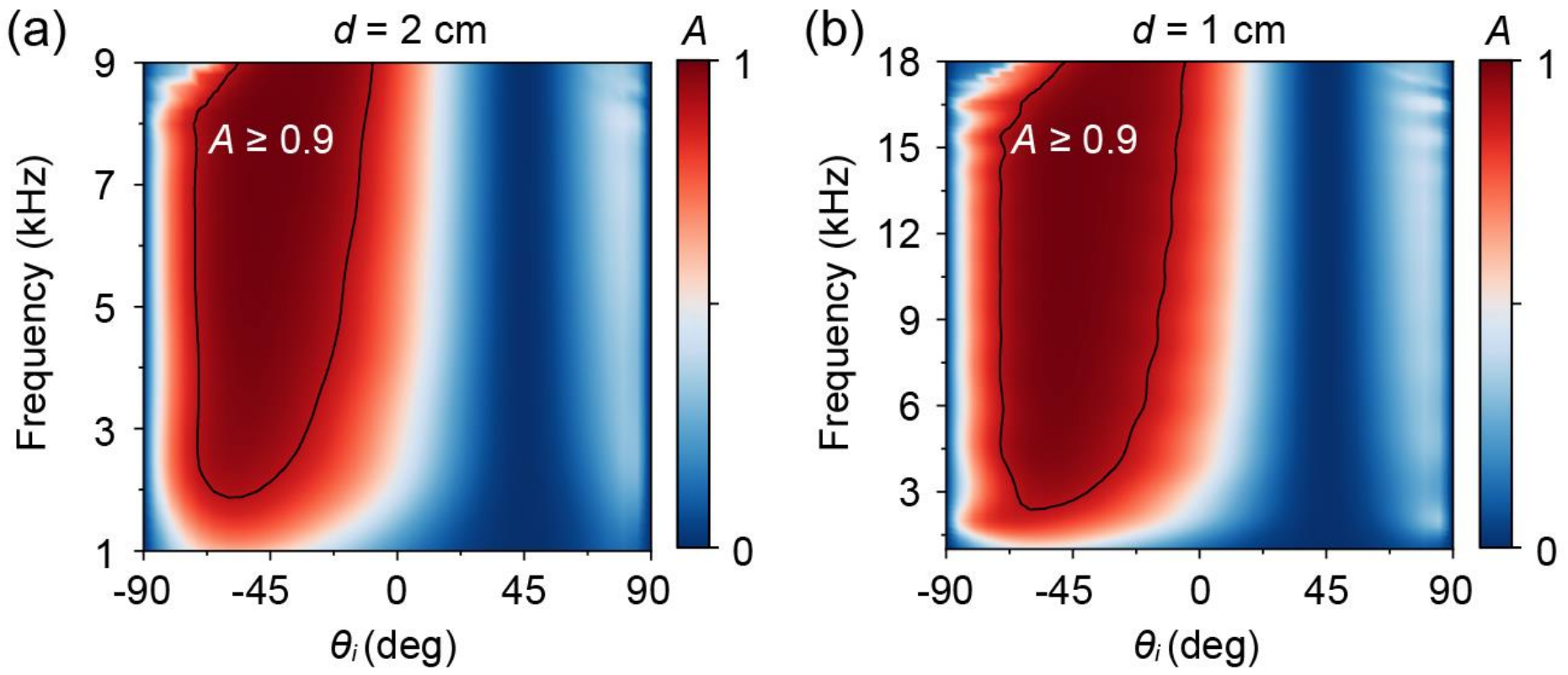

**Fig. S4.** Discussion on the diffractions of the meta-MPPs. (a, b) Simulated absorptance of the meta-MPP regarding the incident angle and working frequency when $d = 2\ cm$ (a) and $d = 1\ cm$ (b).

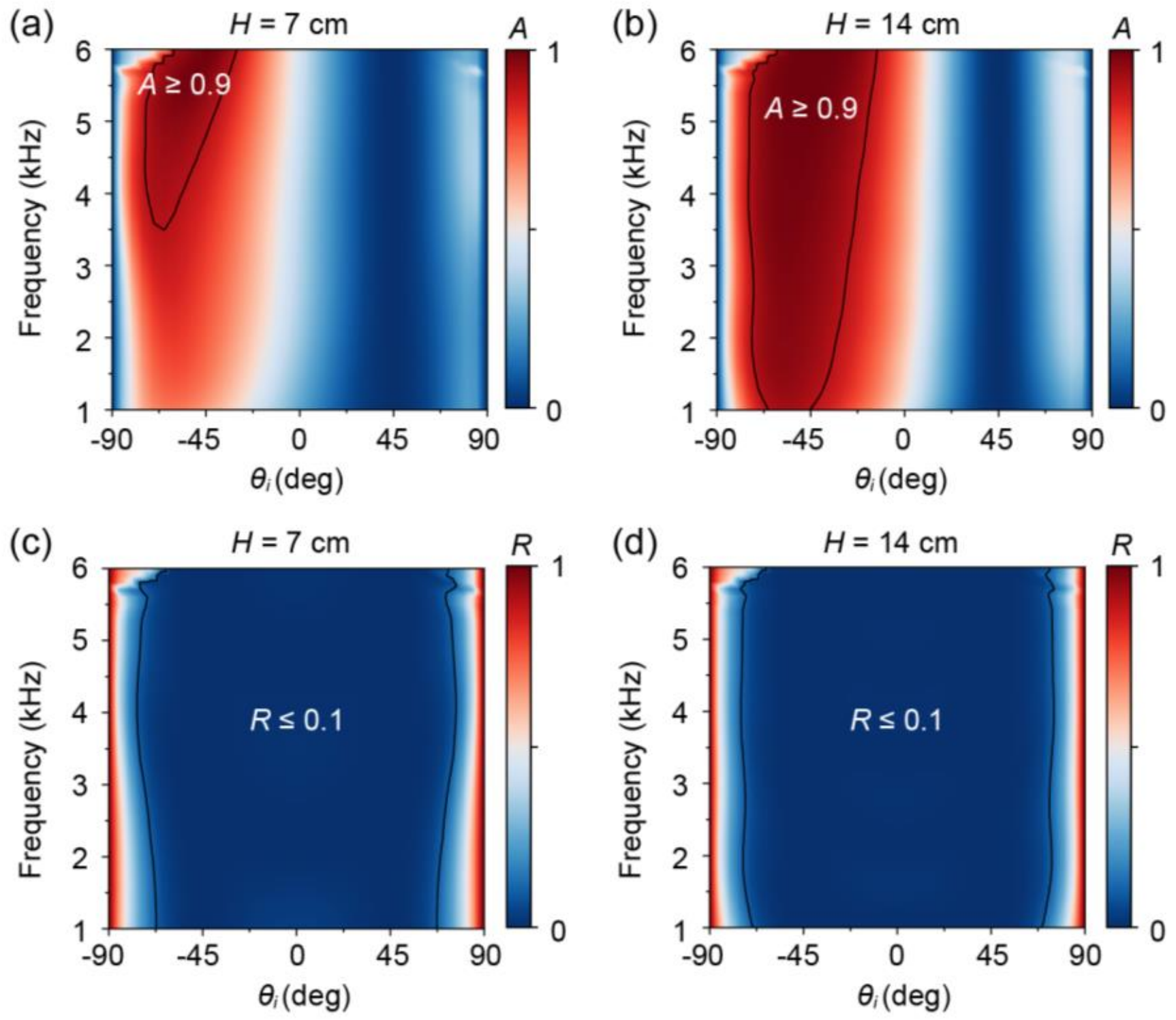

**Fig. S5.** Robustness of the meta-MPPs. (a, b) Simulated absorptance of the meta-MPP regarding the incident angle and working frequency when $H = 7\ cm$ (a) and $H = 14\ cm$ (b). (c, d) Simulated reflectance of the meta-MPP regarding the incident angle and working frequency when $H = 7\ cm$ (c) and $H = 14\ cm$ (d).

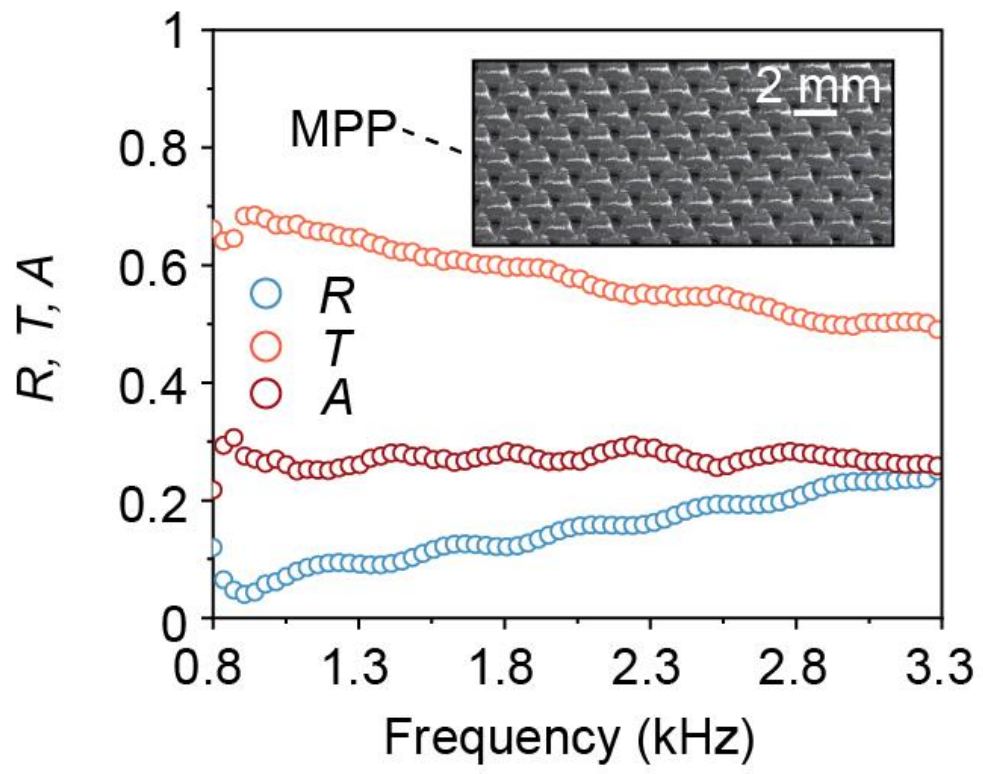

**Fig. S6.** Measured reflection, transmission, and absorption spectra of the MPP under normal incidence. Here, we utilize a type of steel MPP with periodically arranged triangular cone holes, as depicted in the inset of Fig. S6. The measured reflection, transmission, and absorption spectra of the MPP by using the impedance tube method are shown in Fig. S6. It can be seen that the absorptance of the MPP is lower than 0.3 across the working frequencies.

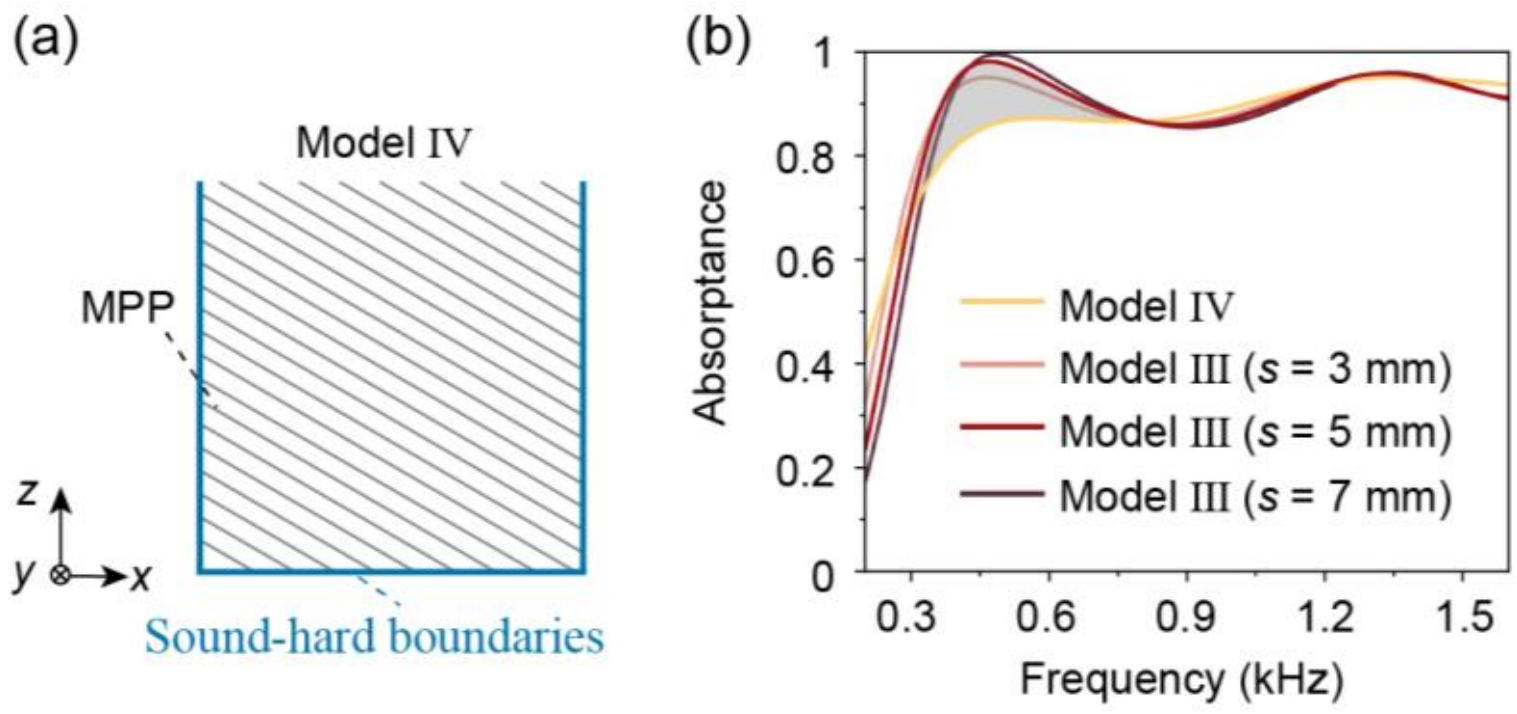

**Fig. S7.** Simulated absorptance curves of model III for different sizes of an opening, *s*. (a) Schematic diagram of model IV. (b) Absorption spectra of model III with different sizes of an opening.

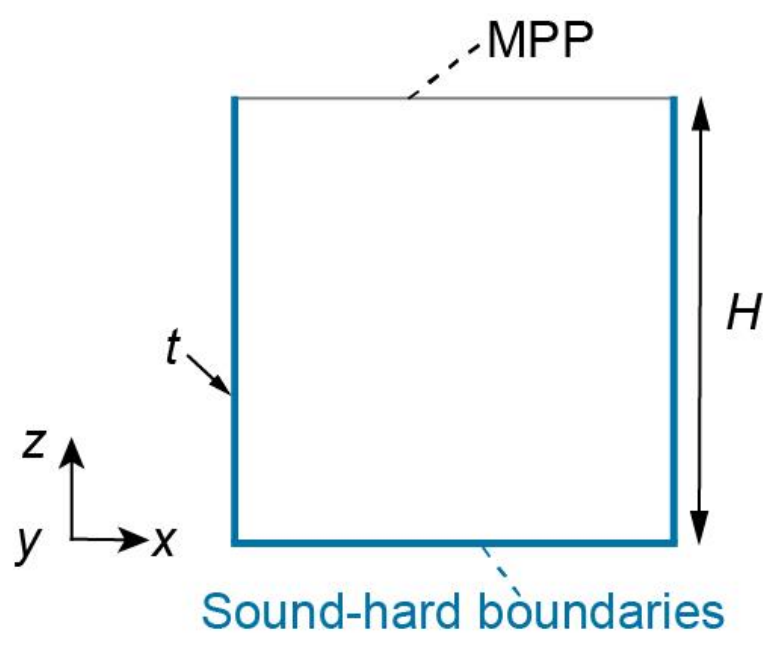

**Fig. S8.** Schematic diagram of the MPP with a cavity. An MPP with a backing cavity is the traditional design method for an MPP sound absorber. The cavity has a thickness of 9.8 cm, and the rigid wall has a thickness of $t = 2\ mm$. Therefore, the total thickness is $H = 10\ cm$.

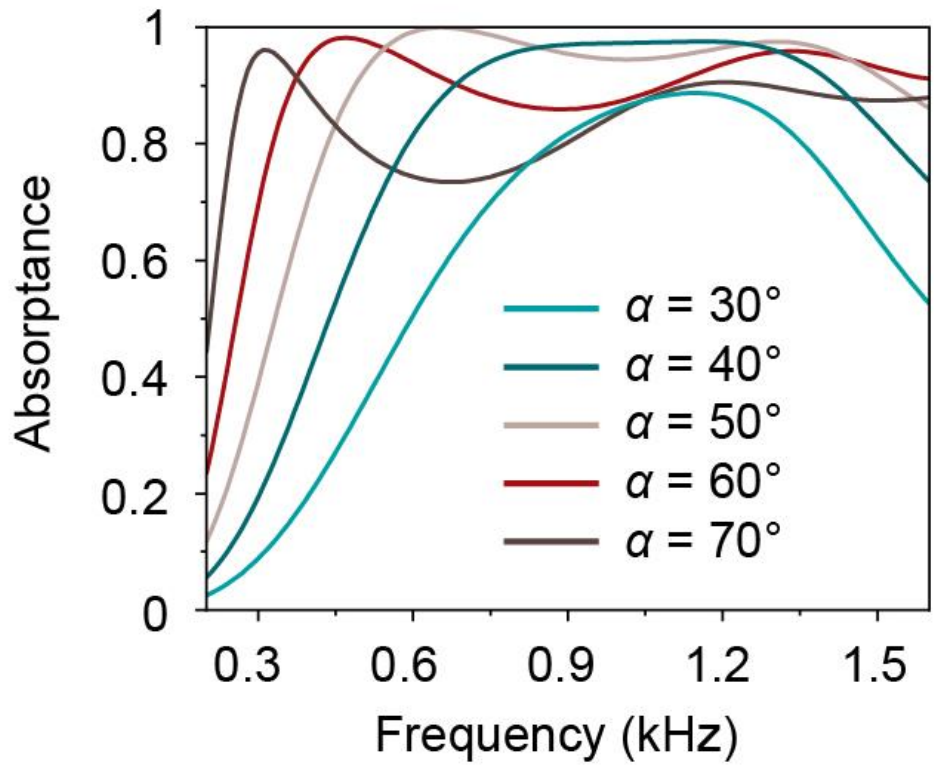

**Fig. S9.** Simulated absorption curves of the sound absorber for varying titled angles of the MPPs. We calculate the absorption curves of the sound absorber for varying titled angles of the MPP for normal incidence. The sound absorber with $\alpha = 60°$ demonstrates superior absorption performance compared to models with $\alpha = 30°, 40°, 50°$. Although the sound absorber with $\alpha = 70°$ shows some improvement in absorption below 0.37 kHz compared to $\alpha = 60°$, it notably decreases above this frequency. Consequently, the model with $\alpha = 60°$ is deemed the optimal choice.

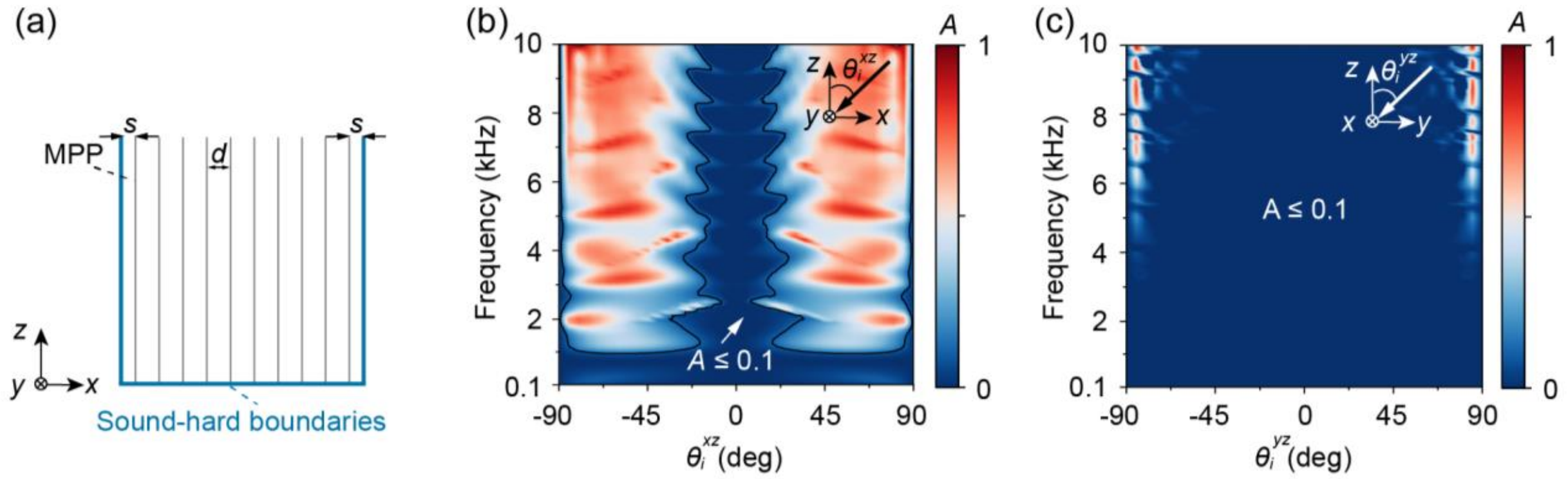

**Fig. S10.** Simulated absorption performance of the absorber when all MPPs are arranged vertically. (a) Schematic diagram of the absorber in the *x*-*z* plane. (b, c) Absorptance of the designed absorber as a function of the incident angle $\theta_i^{xz}$ (b) and $\theta_i^{yz}$ (c) and working frequency.

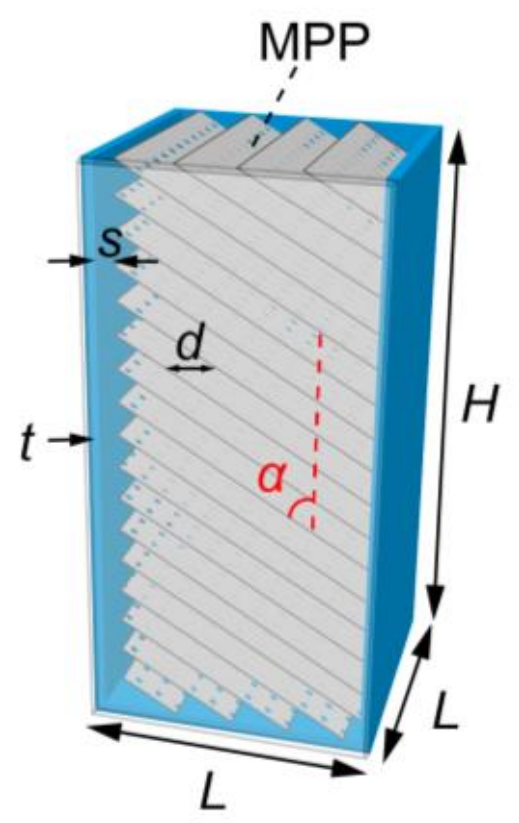

**Fig. S11.** 3D illustration of sample II. Here, relevant parameters are $\alpha = 60°$, $d = 1\ cm$, $s = 3\ mm$, $t = 2\ mm$, $L = 5\ cm$, and $H = 10\ cm$. The parameters of the MPP are the same as that of sample I in the main text.

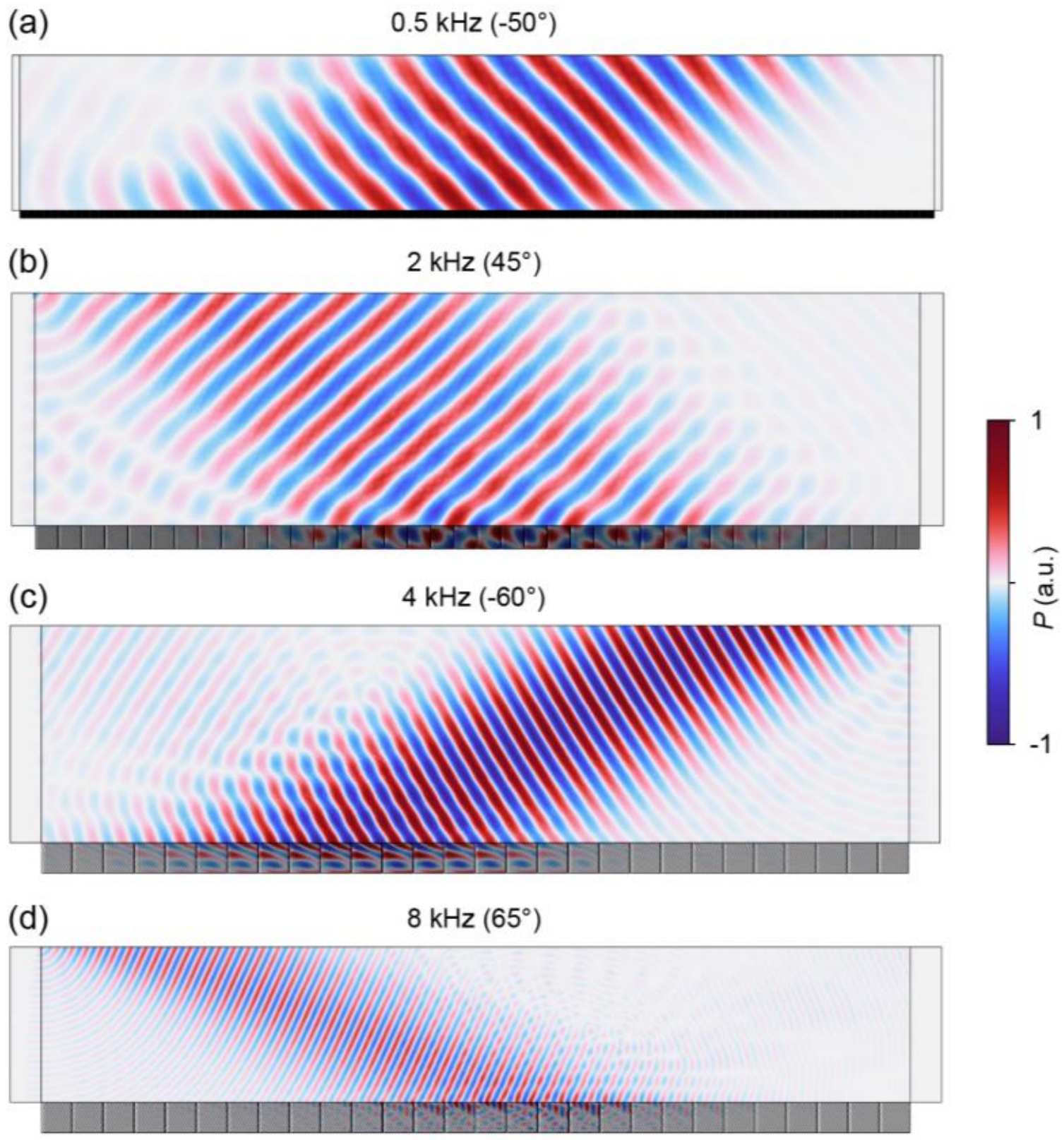

**Fig. S12.** Simulated total pressure distributions of the meta-MPP absorber array at 0.5 kHz (-50°), 2 kHz (45°), 4 kHz (-60°), 8kHz (65°), respectively. Clearly, the meta-MPP absorber exhibits ultrabroadband and omnidirectional absorption capabilities.

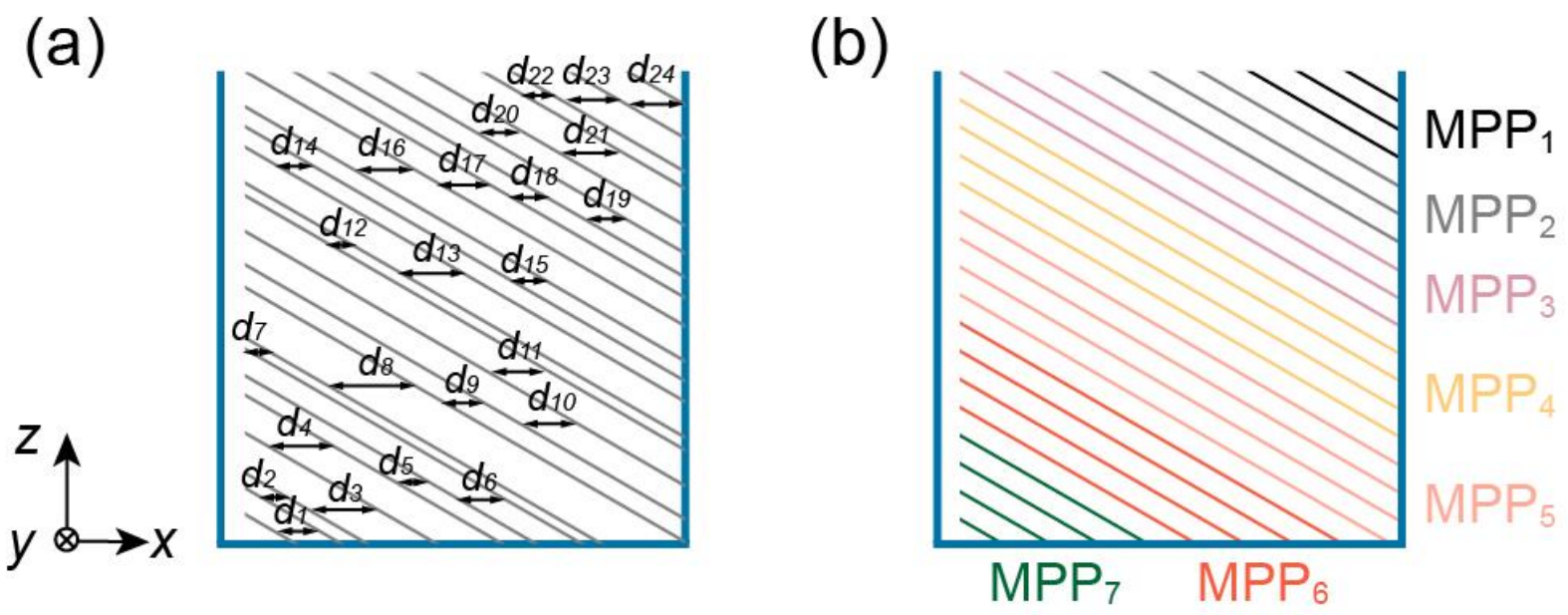

**Fig. S13.** Detailed parameters of the meta-MPP absorbers with random $d$ distribution and random MPP parameters. (a) Relevant parameters are $d_1 = 10\,mm$, $d_2 = 5\,mm$, $d_3 = 15\,mm$, $d_4 = 13\,mm$, $d_5 = 3\,mm$, $d_6 = 10\,mm$, $d_7 = 4\,mm$, $d_8 = 18\,mm$, $d_9 = 8\,mm$, $d_{10} = 12\,mm$, $d_{11} = 13\,mm$, $d_{12} = 5\,mm$, $d_{13} = 13\,mm$, $d_{14} = 7\,mm$, $d_{15} = 8\,mm$, $d_{16} = 12\,mm$, $d_{17} = 10\,mm$, $d_{18} = 8\,mm$, $d_{19} = 9\,mm$, $d_{20} = 9\,mm$, $d_{21} = 14\,mm$, $d_{22} = 6\,mm$, $d_{23} = 12\,mm$, and $d_{24} = 12\,mm$. (b) Relevant parameters are $d_p = 0.18\,mm$, $t_p = 0.15\,mm$, $\sigma = 0.03$ for $MPP_1$, $d_p = 0.2\,mm$, $t_p = 0.1\,mm$, $\sigma = 0.02$ for $MPP_2$, $d_p = 0.21\,mm$, $t_p = 0.12\,mm$, $\sigma = 0.04$ for $MPP_3$, $d_p = 0.19\,mm$, $t_p = 0.11\,mm$, $\sigma = 0.03$ for $MPP_4$, $d_p = 0.25\,mm$, $t_p = 0.1\,mm$, $\sigma = 0.025$ for $MPP_5$, $d_p = 0.17\,mm$, $t_p = 0.15\,mm$, $\sigma = 0.035$ for $MPP_6$, $d_p = 0.15\,mm$, $t_p = 0.1\,mm$, $\sigma = 0.05$ for $MPP_7$.